# An Investigation of Libration Heating and the Thermal State of Enceladus's Ice Shell

Wencheng D. Shao[a, *], Francis Nimmo[a]

[a]*University of California Santa Cruz, 1156 High St., Santa Cruz, CA 95064, the United States*
*Corresponding to: wshao7@ucsc.edu*

**Highlights:**
- We investigate the effect of libration heating on the thermal state of Enceladus's ice shell
- Libration heating in the ice shell is insufficient to match the observed high heat loss on Enceladus
- If in thermal equilibrium, Enceladus is in a stable thermal state resisting small perturbations to the shell thickness
- Thermal runaway or episodic heating on Enceladus is unlikely to originate from librations of the ice shell

**Abstract**

Tidal dissipation is thought to be responsible for the observed high heat loss on Enceladus. Forced librations can enhance tidal dissipation in the ice shell, but how such librations affect the thermal state of Enceladus has not been investigated. Here we investigate the heating effect of forced librations using the model of Van Hoolst et al. (2013), which includes the elasticity of the ice shell. We find that libration heating in the ice shell is insufficient to match the inferred conductive heat loss of Enceladus. This suggests that either Enceladus is not in a thermal steady state, or additional heating mechanisms beneath the ice shell are contributing the bulk of the power. In the presence of such an additional heat source, Enceladus resides in a stable thermal equilibrium, resisting small perturbations to the shell thickness. Our results do not support the occurrence of a runaway melting process proposed by Luan and Goldreich (2017). In our study, the strong dependence of conductive loss on shell thickness stabilizes the thermal state of Enceladus's ice shell. Our study implies that thermal runaway (if it occurred) or episodic heating on Enceladus is unlikely to originate from librations of the ice shell.

1. **Introduction**

Enceladus is the third geologically active body whose internal heat can be detected by remote sensing (Spencer et al. 2006), following the Earth and Io. A high heat flow, ~5-15 GW, is observed by Cassini's Composite Infrared Spectrometer (CIRS) on the south polar terrain (SPT) of Enceladus (Spencer et al. 2006, 2013; Howett et al. 2011). Gravity, shape and libration data from Cassini demonstrate the existence of a global subsurface ocean (e.g., Iess et al. 2014; Beuthe et al. 2016; Thomas et al. 2016; Čadek et al. 2016). If the freezing point of the ocean is close to 273 K, the implied conductive heat flow is a few tens of GW at present (Hemingway et al. 2018). Thus, if Enceladus is now in steady state and the ocean is not freezing with time, a corresponding endogenic heat production must be taking place.

Because radiogenic heating at Enceladus is only ~0.3 GW (Roberts and Nimmo 2008), by far the most likely heat source is tidal dissipation. Until recently, it was unclear whether sufficient tidal heat was available. However, recent astrometric results (e.g., Lainey et al. 2020) have demonstrated that Saturn is transferring energy and momentum to its orbiting satellites at a much higher rate than previously thought (Meyer and Wisdom 2007). Thus, even with the current small eccentricity, tidal dissipation can be sufficient to maintain Enceladus in a conductively steady state (Fuller et al. 2016; Nimmo et al. 2018).

However, several puzzles remain. One is where in Enceladus the tidal heat is dissipated. Initial studies focused on the ice shell (e.g., Roberts and Nimmo 2008; Shoji et al. 2013; Běhounková et al. 2015), but more recent studies have found it difficult to generate enough heat in the ice shell (Souček et al. 2016; Beuthe 2019). Instead, water-filled fractures (Kite and Rubin 2016), the ocean beneath (Chen and Nimmo 2011; Tyler 2011; Wilson and Kerswell 2018; Rovira-Navarro et al. 2019, 2020), or the porous silicate core (Roberts 2015; Choblet et al. 2017; Liao et al. 2020) have been suggested as alternatives. A second is whether Enceladus is actually in steady state. In principle, tidal-orbital feedbacks can arise and generate time-variable heating rates and eccentricities (e.g., Ojakangas and Stevenson 1986). Furthermore, the presence of deformed terrains with indications of high heat flux elsewhere on Enceladus (Giese et al. 2008) suggests some kind of time-variable behaviour.

We will examine both of these questions through the prism of ice shell librations.

In a reference frame fixed to the surface of Enceladus, librations result in a periodic motion of Saturn across the sky. Librations can be either physical - variations in the ice shell rotation rate, driven by time-variable torques from the primary – or optical – driven by the time-variable orbital speed of Enceladus in its eccentric orbit (Hurford et al. 2009). Either kind results in motion of the tidal bulge relative to the solid surface and can thus produce deformation and heating. In general, librations are forced at various distinct periods arising from the various orbital periodicities. However, Enceladus also has free libration periods, the natural frequencies of the system. In the case of an ice shell, the natural frequency depends on the ice shell thickness. As with any oscillator, if the natural libration frequency is close to the forcing frequency, resonant amplification will occur.

An early suggestion that Enceladus's high heating rate was driven by a 3:1 libration resonance (Wisdom 2004) was not borne out by subsequent Cassini gravity measurements. Rambaux et al. (2011) investigated forced librations at Europa and argued that a similar resonance might occur, with the very interesting possibility of a thermal runaway. However, this paper did not include the effect of a shell of *finite* rigidity on the libration amplitude (cf. Van Hoolst et al. 2013). Luan and Goldreich (2017) proposed a thermal runaway for Enceladus. Eccentricity growth of Enceladus produces more tidal dissipation and melts the ice shell. The thinner ice shell in turn leads to enhancement of tidal dissipation. Runaway melting follows until the decrease of shell thickness leads to large eccentricity damping. One goal of our study is to investigate whether a libration-driven thermal runaway could take place at Enceladus or whether thermal equilibrium of the ice shell can be maintained against small perturbations.

Rambaux et al. (2010) investigated forced librations at Enceladus, but their results did not include the effect of a subsurface ocean. As shown by Van Hoolst et al. (2013), inclusion of an ocean overlain by an elastic shell can completely change the librational response of a body. Van Hoolst et al. (2016) included this effect in their study of Enceladus's librations but did not address the issue of heat generation. This is because the present-day libration amplitude is known (Thomas et al. 2016) and is too small to generate sufficient heat. However, since Enceladus's eccentricity and libration amplitudes could have been higher in the past, we wish to investigate whether a thermal runaway could have operated recently, so that it is contributing to the present-day energy budget.

In this work we calculate the forced librations of Enceladus's ice shell using the model of Van Hoolst et al. (2013). We compare the resulting ice shell tidal dissipation rate to the present-day heat flow on Enceladus to understand the thermal state of Enceladus. We find, in common with other studies, that the ice shell heating is insufficient to compensate for the high conductive heat loss on Enceladus. We then investigate whether a thermal runaway of the kind proposed above could occur on Enceladus. We find that even with a higher eccentricity in the past, Enceladus could have been in a stable, high-heat-flux equilibrium resistant to small perturbations, and that no thermal runaway is likely.

This paper is structured as follows. Section 2 describes the methodology and data we use to calculate the forced libration amplitude of Enceladus. Sections 3 and 4 present the forced librations at different frequencies. Section 5 shows the librational heating. Section 6 discusses the possible thermal states of Enceladus. Section 7 summarizes the main results of this work.

## 2. Methodology

We follow the elastic libration model established by Van Hoolst et al. (2013, hereafter VB13) to calculate the forced libration amplitude. This model is developed for a tidally locked satellite with three homogeneous layers: an ice shell, a subsurface ocean and a rocky core. Compared to previous libration models assuming infinite rigidity of the ice shell (e.g., Van Hoolst et al. 2009; Rambaux et al. 2011), this model includes the effect of the finite elasticity of the shell.

In this libration model, the gravitational torques the satellite experiences can be divided into two major parts. One is the total torque applied by the external gravitational potential on both the periodic and static bulges

$$\Gamma_t = \frac{3}{2}\frac{k_f - k_2}{k_f}(B - A)\frac{GM_p}{d^3}\sin 2\psi, \qquad (E1)$$

where $k_f$ and $k_2$ are the fluid Love number and classical dynamical Love number. $(B - A)$ is the equatorial moment-of-inertia difference. $G$ is the gravitational constant, $M_p$ is the mass of the primary, and $d$ is the distance between the satellite and the primary. $\psi$ is the angle between the long axis of the satellite and the direction to the primary. This external torque expression is from equation (12 or 30) in VB13 and is applicable for each internal layer of the satellite. To get an expression for a specific layer, Love numbers, moment-of-inertia difference and the angle $\psi$ need to be specified for that layer. Note that if the satellite has no rigidity, then $k_2=k_f$ and the total torque is zero.

The other part is the torques between internal layers. The pressure torque from the subsurface ocean can be divided into two components which can be incorporated into the expressions of the external and internal gravitational torques. The combined effect of the external gravitational torque above and the contribution from the oceanic pressure torque is

$$\Gamma_{pe,i} = \frac{3}{2}\left[(B_i - A_i)\frac{k_f^i - k_2^i}{k_f^i} + (B_{ob} - A_{ob})\frac{k_f^{ob} - k_2^{ob}}{k_f^{ob}}\right]\frac{GM_p}{d^3}\sin 2\psi_i, \qquad (E2)$$

for the rocky core. Subscript or superscript $i$ represents the core (or the rocky interior), and $ob$ refers to the bottom of the ocean. This expression is from equation (34) in VB13. A similar expression can be written for the ice shell layer, with subscript or superscript substituted.

The other component of the oceanic pressure torque goes into the expression of the internal gravitational torque. The final expression of the internal torque including the contribution from the pressure torque and associated with the static bulge is

$$\Gamma_{pg,i}^{static} = \frac{4\pi G}{5}[(B_i - A_i) + (B_{ob} - A_{ob})][\rho_o\beta_o + \rho_s(\beta_s - \beta_o)]\sin 2(\gamma_s - \gamma_i), (E3)$$

where subscript or superscript $o$ and $s$ represent the ocean layer and the shell layer, respectively. $\rho$ and $\beta$ are the density and the equatorial flattening. $\gamma$ is the small libration angle. This expression is from Eq. (37) in VB13. This expression is the internal torque exerted on the core, and the same torque with an opposite direction is exerted on the shell. The internal torque associated with the periodic bulge and including the contribution from the pressure torque is

$$\Gamma_{pg,i}^{per} = 2K_{im}L_{osc} + 2K_{ii}\gamma_i - 2K_{is}\gamma_s, \quad (E4)$$

for the core layer. Here the coefficients $K_{im}$, $K_{ii}$ and $K_{is}$ refer to equation (39-41) in VB13. $L_{osc}$ is the oscillation part of the true longitude at the diurnal frequency (refer to equation (11) in Rambaux et al. 2011). This expression is from equation (38) in VB13. A torque equal to (E4) but with an opposite direction is exerted on the shell layer. Note that here we only consider the internal torque related to the periodic tidal bulge at the diurnal frequency. The torque related to the periodic bulge at long-term frequency is not considered in this study.

The deformation of each layer also changes the polar moment of inertia. The change of the polar moment of inertia due to the variable centrifugal acceleration has negligible effect (Van Hoolst et al. 2008, 2013) and is thus not considered here. The change due to zonal tides at the diurnal frequency gives an additional forcing of a few percent (Van Hoolst et al. 2008, 2013) and is included in our calculations. Refer to equation (44) in VB13 for the expression of this effect. Here we ignore the effect on the polar moment of inertia from zonal tides at long-term periods.

The final equations to calculate the forced librations are

$$C_s\ddot{\gamma}_s + K_1\gamma_s + K_2\gamma_i = 2K_3L_{osc}, \quad (E5)$$
$$C_i\ddot{\gamma}_i + K_4\gamma_s + K_5\gamma_i = 2K_6L_{osc}, \quad (E6)$$

where $C$ is the polar moment of inertia. The expressions of the coefficients $K_1$ to $K_6$ refer to Equations (47-52) in VB13. To allow readers to verify our calculations, an example of this calculation is shown in Appendix B.

In the libration calculations, the dynamical Love number $k_2^j$ for each layer needs to be specified. This Love number can be acquired from the radial displacement:

$$k_2^j = \frac{4\pi G\rho_j}{5R^3}\left(R_j^4 y^j - R_{j-1}^4 y^{j-1}\right), \quad (E7)$$

where $R$ is the radius of the satellite. $R_j$ is the radius of the upper surface of layer $j$. $j-1$ refers to the layer below layer $j$. $y^j$ is the radial displacement at the upper surface of layer $j$. This expression is from equation (24) in VB13. Here we calculate the radial displacement using the viscoelastic model developed by Roberts and Nimmo (2008). The model solves the correspondence principle to get the radial displacement. The input of this model is the physical properties of each homogeneous layer, including rigidity and viscosity. In our work, to better represent the viscosity profile in the ice shell, we divide the shell region into multiple layers with different viscosity values. The details of this division are described in Section 4.

The true-longitude oscillation ($L_{osc}$) is obtained from the JPL/Horizon database (https://ssd.jpl.nasa.gov/horizons.cgi). This database conveniently provides the time series of orbital osculating elements for different astronomical bodies. Here we obtain the time series of the orbital elements of Enceladus for 300 years (1850-2150) with a time interval of one hour. The reference frame is ICRF/J2000.0, and the center is Saturn. We detrend the calculated true-longitude data to get its oscillations. Then through Fourier decomposition, the orbital forcings at different frequencies are obtained (see Table 1). Finally, the libration amplitudes are calculated via (E5-E6), and the libration heating is calculated via (E8) in Section 5.

## 3. Forcings in JPL/Horizon data

Table 1 Fourier decomposition of true-longitude oscillations of Enceladus based on data from JPL/Horizons Ephemeris.

| Frequency (rad/day) | Period (days) | Magnitude (arc second) | Phase (degree) |
|---|---|---|---|
| 4.579656 | 1.371978 | 1671.31 | 111.64 |
| 0.001548 | 4058.259259 | 901.51 | 163.83 |
| 0.004415 | 1423.025974 | 623.74 | 46.61 |
| 4.578107 | 1.372442 | 36.39 | -112.19 |

Table 1 gives the main orbital perturbing terms of Enceladus, obtained from a frequency analysis of true-longitude oscillations via the fast Fourier transform (FFT) method. The first three forcing terms are the mean anomaly, the Dione-Enceladus resonance and the Dione proper pericenter, respectively (Rambaux et al. 2010). The fourth, small term is FFT spectral leakage from the mean anomaly.

Comparing this table to that in Rambaux et al. (2010), here we obtain a slightly smaller amplitude for the mean-anomaly forcing. This is due to spectral leakage of the FFT method. Minimizing this leakage effect is not conducted here since it does not significantly affect our results. The phases of the first three terms here are different from those in Rambaux et al. (2010) because of difference in selected datasets (e.g., different initial starting points of time series of orbital elements). This phase difference does not affect accuracy of our calculations; in Figure S1, we show that our dataset and method give results consistent with Figure 2a in Rambaux et al. (2010) if the same interior model is used.

## 4. Interior models and librational response

We first need to specify the interior structure of Enceladus to calculate forced librations from equations (E5-E6). We use a three-layer model here: core, ocean and shell. Based on Cassini's measurement of the mean moment of inertia (0.331±0.002), the core size of Enceladus is inferred as ~190 km (Hemingway et al. 2018), assuming Enceladus has fully differentiated. We adopt this core size as a constraint. For the ice shell thickness, various estimates give a range of results based on different methods (Iess et al. 2014; McKinnon 2015; Beuthe et al. 2016; Thomas et al. 2016; Čadek et al. 2016; Van Hoolst et al. 2016; Hemingway and Mittal 2019). Therefore, here we take the shell thickness as a free parameter, ranging from 5 to 50 km, and construct 41 interior models with different shell thicknesses.

Table 2 shows major physical properties of our interior models. Bulk modulus, shear modulus and viscosity are required to calculate the radial displacement (equation E7) in the tidal model of Roberts and Nimmo (2008). The bulk modulus (not shown in Table 2) is set as $10^{19}$ Pa for all layers (effectively incompressible). The shear modulus assumed for the core is $10^{10}$ Pa and for the ice shell is $3.3 \times 10^9$ Pa. The core viscosity is $10^{25}$ Pa s. Because the code of Roberts and Nimmo (2008) cannot treat a purely fluid layer, the ocean is represented as a layer with low viscosity and rigidity. As long as this layer's Maxwell time is well away from the forcing period, this approximation works reasonably well. To better account for the viscosity effect of the ice shell, here we divide the shell into multiple sub-layers with viscosity changing by orders of magnitude. Details of this treatment are described in Appendix A.

**Table 2** Physical properties of interior models of Enceladus

|  | Core | Ocean | Shell |
|---|---|---|---|
| Density (kg/m$^3$) | Calculated[a] | 1000 | 900 |
| Radius of upper boundary (km) | 190 [b] | 202.3-247.3 | 252.3 |
| Shear modulus (Pa) | $10^{10}$ | $10^6$ | $3.3\times10^9$ |
| Viscosity (Pa s) | $10^{25}$ | $10^8$ | $10^{14}$ [c] |

[a] Core's density is calculated using the constraint of the total mass of Enceladus
[b] See Hemingway et al. (2018)
[c] This value is the basal viscosity of ice shell. Viscosity varies within ice shell; See Appendix A for details

In common with earlier works (e.g., Van Hoolst et al. 2016), our calculations show that the diurnal libration amplitude is greatly dependent on the shell thickness (Figure 1a). The diurnal libration amplitude increases from a few hundred meters to a few kilometers as the shell thickness decreases from 50 km to 5 km. This is because one of the two free frequencies gets closer to the diurnal frequency as the shell becomes thinner, which amplifies the diurnal libration. On the contrary, the libration amplitudes at long periods, 1423 days and 4058 days, are almost constant for different interior models. This is because when the interior structure changes, the free-libration periods change but are still around a few days. The long forcing periods are always far away from the free libration periods. This insensitivity has also been seen in the long-term librations of Europa (Rambaux et al. 2011).

Despite the insensitivity to the shell thickness, the long-term libration amplitudes are not negligible (~0.7 and ~1.1 km) compared to the diurnal libration amplitude. When observations are being interpreted, it is important to disentangle the diurnal and long-term librations. Once the diurnal libration is extracted from observations, then utilizing the sensitivity of this libration, the shell thickness can be constrained. This was done by Van Hoolst et al. (2016), who used the libration data from Thomas et al. (2016) and got an average shell thickness of 14-26 km for Enceladus. Using our model, we infer a shell thickness of 15-19 km corresponding to the libration data of Thomas et al. (2016) (Figure 1a).

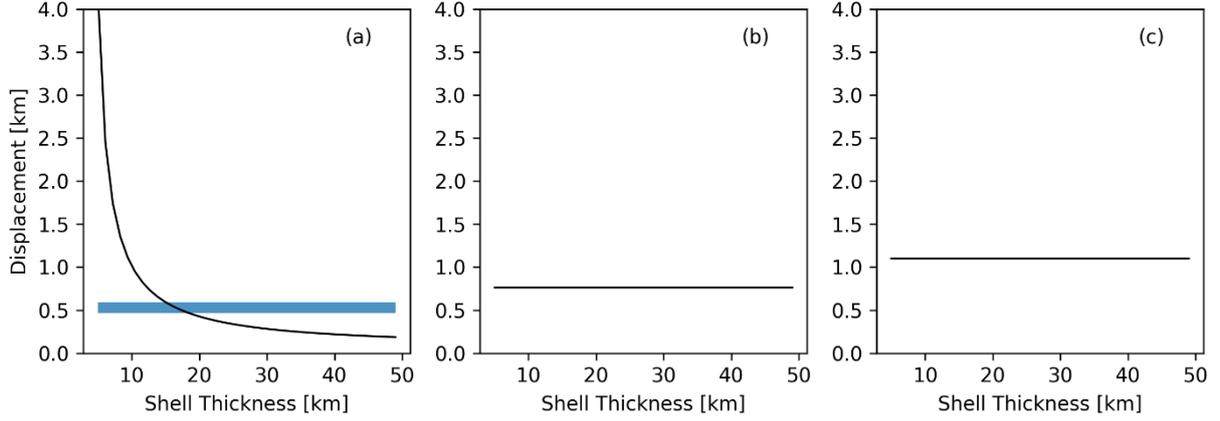

**Figure 1** Dependence of libration amplitudes (or displacements) at (a) 1.37-day, (b) 1423-day and (c) 4058-day periods on shell thickness. Blue region gives the observed libration amplitude at equator (Thomas et al. 2016).

## 5. Diurnal libration heating

Here we consider the heating effect of the diurnal forced libration. The heat contributed by long-term forced librations is assumed to be insignificant and not included here. The total tidal dissipation in the ice shell, including the diurnal libration heating effect, can be approximated by a simple equation (Wisdom 2004; Hurford et al. 2009):

$$E_t = \left[\frac{9}{2}e^2 + \frac{3}{2}(2e+F)^2\right] \times \frac{k_2}{Q_s}\frac{GM_p^2 n R_s^5}{a^6}, \quad (E8)$$

where $e$ is eccentricity, and $F$ is libration amplitude. $k_2$ and $Q_s$ are the Love number and the dissipation factor of the satellite. Note that $k_2$ here is for the whole satellite, different from that in equation (E7) for a specific layer. $G$ is the gravitational constant. $M_p$ is the mass of the primary. $a$ is orbital radius. $n$ and $R_s$ are the mean motion and the radius of the satellite. This equation requires that the diurnal forced libration is out-of-phase with the optical libration (Hurford et al. 2009; Tiscareno et al. 2009), which is true for all our cases.

$k_2$ and $Q_s$ here are obtained from the complex degree-2 tidal Love number $\widetilde{k_2}$ calculated by the tidal model of Roberts and Nimmo (2008)

$$k_2 = Real(\widetilde{k_2}), \quad Q_s = \frac{Real(\widetilde{k_2})}{Imag(\widetilde{k_2})}. \quad (E9)$$

The complex Love number $\widetilde{k_2}$ relates to the surface value of the potential $\widetilde{y_5}$ solved by the correspondence principle (Tobie et al. 2005; Roberts and Nimmo 2008)

$$\widetilde{k_2} = -(\widetilde{y_5}|_{r=R_s, l=2}) - 1. \quad (E10)$$

The tilde indicates a complex number, and $l$ is the spherical harmonic degree.

Our calculations show that the total tidal dissipation in the ice shell is dependent on the shell thickness (Figure 2). The total tidal heating in the ice shell increases as the shell becomes thinner. This is due to the increase of both the Love number $k_2$ and the diurnal libration amplitude $F$ (Figure 1a). The dissipation factor $Q_s$ increases (i.e., becomes less dissipative) by a factor of ~2 as the shell thickness decreases from 50 to 5 km, due to the decreased volume of low-viscosity ice region. The combined effect of the three factors, $k_2$, $F$ and $Q_s$, gives the dependency of the tidal heating in Figure 2a: from 50 km to 10 km for the shell thickness, the heating increases slowly; from 10 km to 5 km, the heating rapidly increases. When the ice shell is ~5 km thick, the tidal heating in the ice shell could reach ~10 GW.

Enhancement of the shell tidal dissipation due to including the diurnal forced libration increases as the shell gets thinner (Figure 2a). When the shell is ~17 km thick (within the inferred shell thickness range from our model and the libration observation), the diurnal forced libration increases the shell heating by ~27%. This percentage is consistent with that in Beuthe (2019), ~28% (the small difference of this value comes from the small difference between our calculated libration and the libration value used in that paper). When the shell is very thin, the heating enhancement due to the forced libration dominates over that due to regular eccentricity tides.

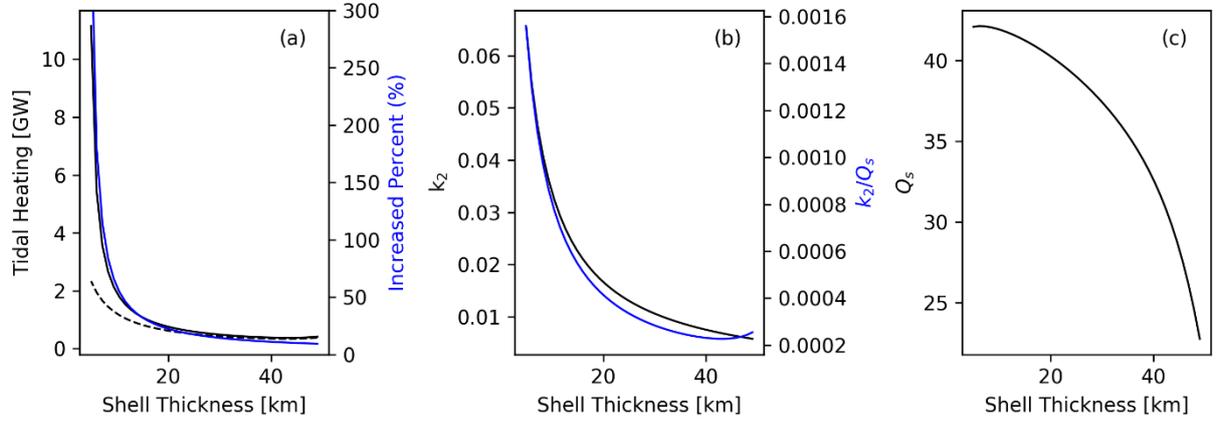

**Figure 2** (a) Shell tidal dissipation rate with (black solid line) and without (black dashed line) the diurnal forced libration included. Blue line shows the increased percentage of shell dissipation by including the diurnal forced libration. (b) Love number $k_2$ for different interior models. (c) Dissipation factor $Q_s$ for different interior models. The ratio of $k_2$ to $Q_s$ is also shown by the blue line in panel b.

We compare the total tidal heating rate to the surface heat flow of Enceladus to understand its current thermal state. Under the assumption that the ice shell is conductive and other assumptions as in Ojakangas and Stevenson (1989), the heat flow can be approximated as

$$H = -S\frac{c}{d}\ln\left(\frac{T_s}{T_b}\right), \qquad (E11)$$

(Hemingway et al. 2018). $S$ is satellite global area. $c$ is a constant (567 W/m, e.g., Klinger 1980). $d$, $T_s$ and $T_b$ are the shell thickness, the surface temperature and the bottom temperature of the ice shell, respectively. The surface temperature is set at 75 K, and the bottom temperature at 273 K. A shell thickness of ~17 km implies a global conductive heat flow of ~34 GW.

However, even with the diurnal libration heating included, the total heating rate in the shell is still insufficient to explain the high heat flow on Enceladus's surface. In the model with the shell thickness of ~17 km, the total shell heating rate is ~0.9 GW (Fig 2a), less than the observed high heat flow on the SPT (~5-15 GW, Spencer et al. 2006, 2013; Howett et al. 2011) and certainly much less than the global conductive heat flow of ~34 GW required to maintain the shell in equilibrium. Only in extreme cases with very thin shells can the total shell heating rate approach 10 GW (Figure 2a). However, these extreme cases also have extreme conductive cooling rates, and the shell's tidal heating in total is still not enough to balance the heat budget. Thus, if Enceladus is currently balancing its heat generation and loss, dissipation must be happening in either the ocean or the silicate interior, as well as in the ice shell.

## 6. Possible thermal equilibrium states of Enceladus

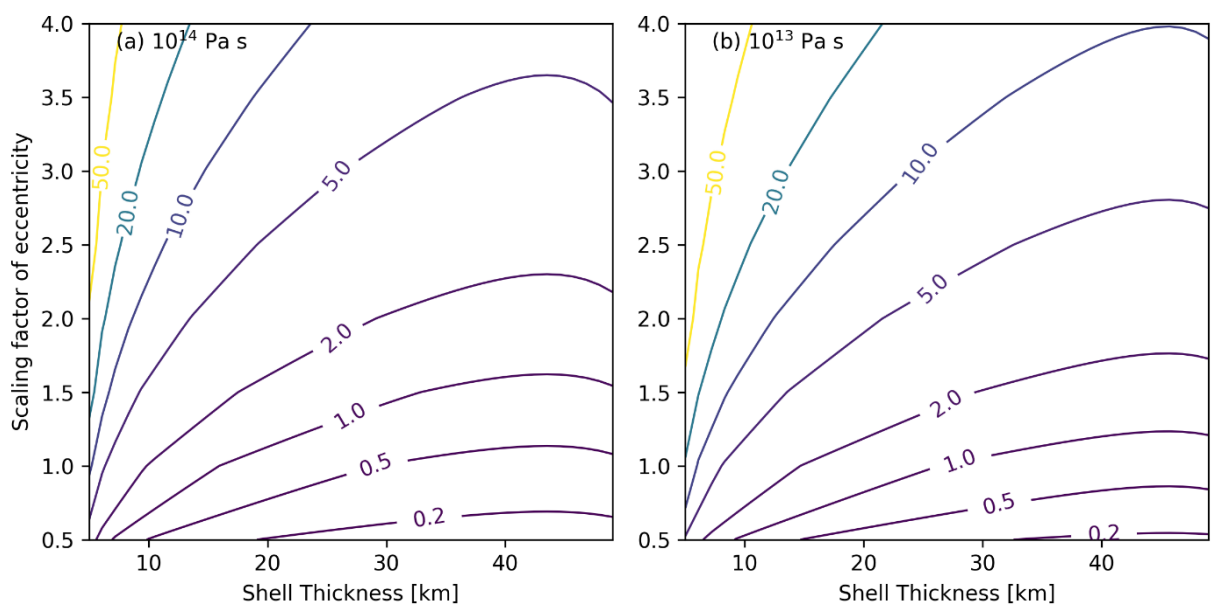

**Figure 3** Dependence of total shell heating rate (in GW) on shell thickness and eccentricity. Value of y axis is the scaling factor with respect to the current eccentricity of Enceladus (0.0047). Bottom viscosity of the ice shell is $10^{14}$ Pa s for (a) and $10^{13}$ Pa s for (b).

To consider Enceladus's possible thermal states in its past, we calculated the shell's total tidal dissipation with different eccentricities (Figure 3). Different shell basal viscosities are also considered. Basically, larger eccentricity and lower viscosity produce more heat in the ice shell. From equation (E8) the shell heating rate is proportional to $e^2$; if the eccentricity was 2 times the current value, the shell heating rate would have been 4 times the current value. As for the basal viscosity, the lower basal viscosity means larger low-viscosity volume in the ice shell and thus more tidal dissipation. For a fixed eccentricity the heat production generally decreases as shell thickness increases, because $k_2/Q$ and the forced libration amplitude both generally decrease with increasing shell thickness (Figure 2).

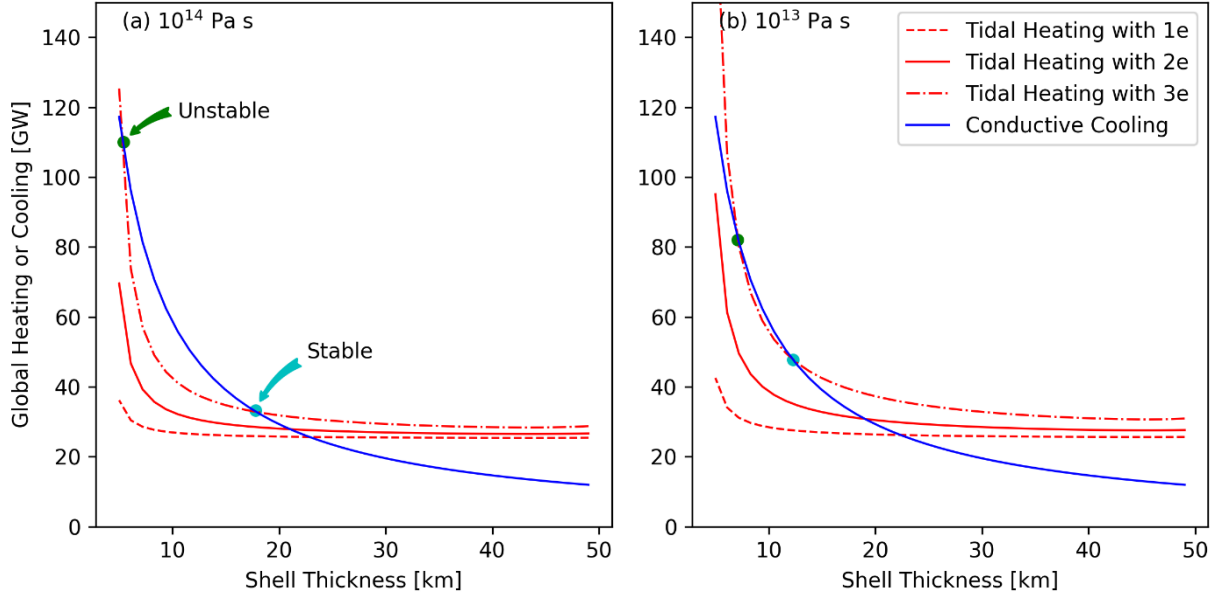

**Figure 4** Total global heating rate (red) versus conductive cooling rate (blue) for interior models with the shell basal viscosity of (a) $10^{14}$ Pa s and (b) $10^{13}$ Pa s. The heating consists of dissipation in the ice shell and a heat source of 25 GW below the shell. Different line styles indicate different orbital eccentricities. Enceladus's surface temperature is taken as 75 K. Stable and unstable equilibrium points are marked out.

Assuming Enceladus's shell is losing heat through conduction, we determine possible thermal equilibria of the shell. As mentioned in previous section, there are possibly other heat sources inside Enceladus beside those in the shell. Perhaps tidal friction in a porous core (Roberts 2015; Choblet et al. 2017; Liao et al. 2020) or turbulent dissipation in the ocean (e.g., Wilson and Kerswell 2018) could generate a significant amount of heat. Here we do not identify other heat sources; instead, we simply use an additional constant heating rate to represent the unknown heating mechanism(s) inside Enceladus. Note that the constant value assumed, 25 GW, is consistent with the astrometrically-derived heating rates (Fuller et al. 2016).

With this additional heating, present-day Enceladus could readily be in a stable thermal-equilibrium state resistant to small perturbations. The total tidal heating of Enceladus's ice shell is relatively insensitive to the shell thickness, inherited from our assumption that the unknown heating mechanism(s) is(are) independent of the shell thickness. In this situation, once thermal equilibrium is reached, a small perturbation to the shell thickness would be resisted by this thermal state. For example, if the shell thickness is decreased by a small percentage, the conductive cooling rate will become larger than the total heating rate (Figure 4). This would cause the ice shell to freeze and return to the original equilibrium point. At the present day, therefore, it appears that heat in the ice shell represents a small fraction of the total heat, and Enceladus could maintain thermal equilibrium against small perturbations to ice shell thickness.

However, it is also possible that Enceladus is not heat-balanced. In this situation, Enceladus is cooling, and its subsurface ocean is freezing with time (i.e., to the left of the stable point in Figure 4). Or on the contrary, Enceladus is heating, and its ice shell is melting (i.e., to the right of the stable point in Figure 4). In either case Enceladus will tend to move back towards equilibrium, but the details would depend on the melting/freezing timescale of the ice shell compared to the eccentricity damping timescale and could in principle result in oscillatory behaviour. Coupled thermal-orbital models of this kind have been investigated in the past (e.g., Ojakangas & Stevenson 1986; Shoji et al. 2014), but are beyond the scope of this work.

Even with a higher eccentricity, Enceladus will still generally be in stable thermal equilibrium (Figure 4). Thus, a thermal runaway of the kind envisaged by Luan and Goldreich (2017) is unlikely to occur in the recent past. In the thermal-runaway scenario, when the ice shell thickness decreases, tidal dissipation increases, and this in turn melts ice and decreases the ice shell thickness further. Runaway melting may occur. Our results, however, do not support this scenario. In our work, equation (E11) gives a strong dependence of conductive loss on shell thickness. In Figure 4, conductive heat loss increases more rapidly than tidal dissipation as ice shell gets thinner, and this prevents the thermal runaway of Enceladus from occurring. This stable equilibrium implies that the multiple resurfacing events inferred for the past of Enceladus (e.g., Giese et al. 2008) may not have arisen from some intrinsic instability of ice shell. More likely possibillities include passage through earlier orbital mean-motion resonances (e.g. Meyer & Wisdom 2008a) or (perhaps) impactors (Roberts and Stickle 2021).

In some extreme cases (thin ice shell, high eccentricity and low viscosity), there is an unstable equilibrium point vulnerable to small perturbations (Figure 4). If Enceladus was once in this point, a runaway process as described above could have occurred. But, quite apart from the extreme parameter choices required, we are then faced with the question of how to put Enceladus into such an unstable point. More information about Enceladus's past is needed to facilitate the investigation of such an unstable thermal state.

## 7. Discussion and conclusions

In this study, we investigated the libration heating effect on Enceladus's thermal state. We found that the ice shell tidal dissipation including the diurnal libration heating is significantly dependent on ice shell thickness, and a ~17 km thick ice shell can generate heat of ~0.9 GW. When the ice shell is very thin, heat enhancement due to the diurnal forced libration dominates the heat generation in the ice shell. The ice shell dissipation is far from being sufficient to match the conductive cooling rate (~34 GW for a ~17 km thick shell) required for Enceladus to be in steady state. Either present-day Enceladus is not in thermal equilibrium, or there are additional large heating sources beneath the ice shell, keeping Enceladus in steady state. If, as seems likely, these additional heating sources are independent of shell thickness, Enceladus could be in a thermal equilibrium state where small perturbations to shell thickness are resisted.

A higher eccentricity Enceladus in the past would also likely have been in a similar, thermally stable state resistant to small perturbations. Any runaway melting process (if it occurred) or episodic heating is unlikely to have originated from librations of the ice shell. There are unstable equilibrium points under some extreme cases (thin shell, high eccentricity and low viscosity), but many unknowns about Enceladus's history leave the investigation of such an unstable point to the future.

One deficit of our study is that we do not calculate the mutual feedbacks between thermal evolution and orbital evolution. Ojakangas and Stevenson (1986) did this coupling for Io and found that unstable or periodic regime can occur. Even though they focused on the convective heat flow on Io, their results also implied a possible cyclic solution for the resurfacing of Enceladus. Meyer and Wisdom (2008b) found that the Ojakangas and Stevenson (1986) mechanism did not produce periodic behavior at Enceladus, but there are certainly other possible modes of cyclic behavior, including that suggested by Luan and Goldreich (2017). While our results do not suggest that librationally-driven thermal runaways occur at Enceladus, coupling of thermal evolution and orbital evolution is a rich topic, and whether it can solve the question of episodic heating events at Enceladus deserves more attention in the future.

**Acknowledgements**

This work is supported by the China Scholarship Council Fellowship to W. D. S.. We would like to thank two anonymous reviewers for constructive comments on the manuscript.

**Supplementary material**

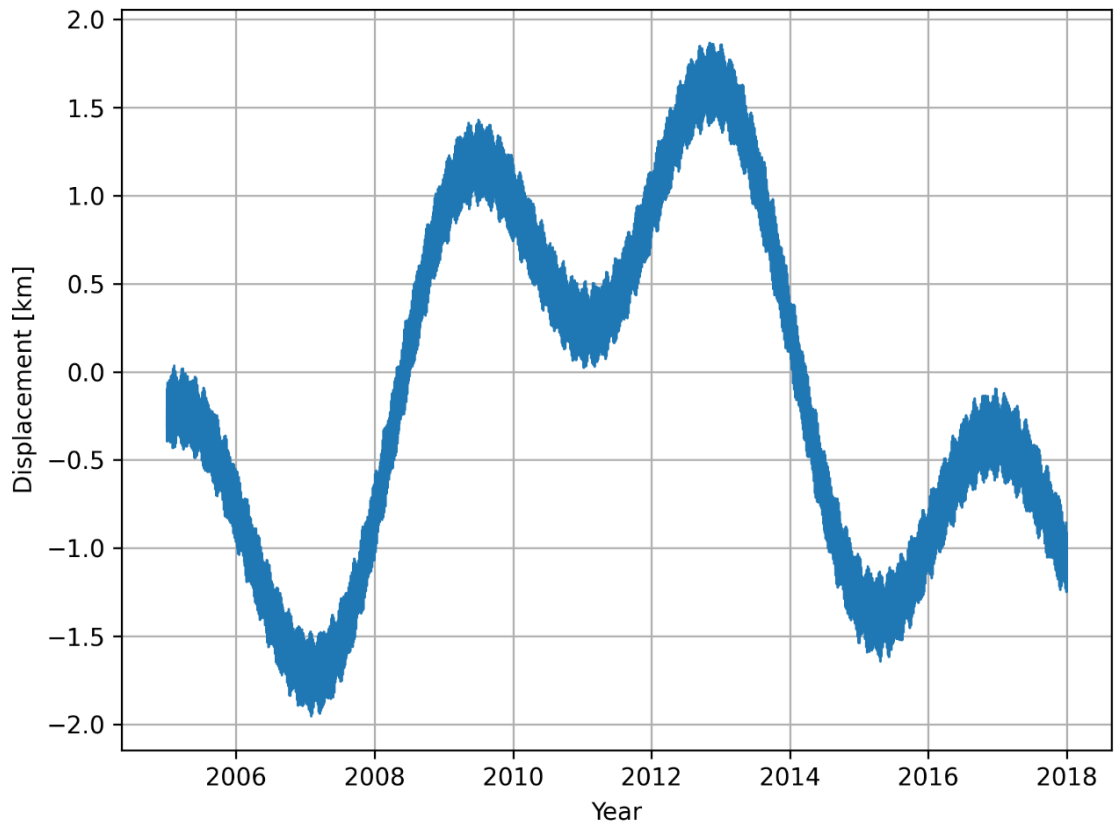

**Figure S1** Variation of physical libration displacement (km) of Enceladus over 2005-2018 period. In this calculation, we used the same interior model as that for Figure 2a in Rambaux et al. (2010). We did not include viscosity effect here since it only gives a small change to the results (refer to Rambaux et al. 2010). We did not include the free libration here since our model may break around the free frequency due to the small amplitude assumption (Rambaux et al. 2011), and libration terms in Rambaux et al. (2010) also did not include the free-frequency term. This figure is aimed to be a reproduction of Figure 2a in Rambaux et al. (2010) to validate the accuracies of both the dataset and the method we used in this work.

## Appendix A. Viscosity profile in the ice shell

The tidal model of Roberts and Nimmo (2008) requires specification of viscosity for each layer. For viscosity of the silicate core, we adopt a large value, $10^{25}$ Pa s (Table 2). For the viscosity of the ocean, we adopt a small value, $10^8$ Pa s (Table 2). For the viscosity of the ice shell, we further divide the shell into multiple sub-layers with different viscosities to more accurately calculate Enceladus's radial displacements and dissipation factor.

According to Ojakangas and Stevenson (1989), under reasonable assumptions, the gradient of $\ln(T)$ is nearly constant through the majority of the ice shell, where $T$ is temperature. Thus, the temperature profile can be approximated as

$$\ln T = \frac{\ln(T_b/T_s)}{d} z + \ln T_s, \quad (A1)$$

where $T_b$ and $T_s$ are the bottom and surface temperature of the ice shell, $d$ is the shell thickness, and $z$ is the vertical coordinate (positive being downward). Ice viscosity $\eta$ relates to temperature through the Arrhenius relation

$$\eta = \eta_0 \exp\left\{l\left(\frac{T_M}{T} - 1\right)\right\}. \quad (A2)$$

Here $T_M$ is the melting temperature of ice, $l$ is a coefficient, and $\eta_0$ is the viscosity at the melting temperature. In this paper, we assume $T_b = T_M = 273\ K$, $l = 24.0$, and $\eta_0 = 10^{14}\ Pa \cdot s$. $\eta_0$ is also the basal viscosity of the ice shell.

We use equations (A1-A2) to construct the sub-layers in the ice shell. We first set the viscosity ranging from $10^{14}$ to $10^{22}$ Pa s (values above $10^{22}$ Pa s will contribute negligible dissipation), then use equation (A2) to derive the temperature, and finally use equation (A1) to derive the ratio $z/d$. This ratio is assumed as the radial position of the midpoint of each sub-layer, and then the positions of the lower and upper surfaces for each sub-layer can be obtained through average of adjacent midpoints. Table A1 shows the calculated results for each sub-layer.

**Table A1.** Viscosity for each sub-layer.

| Sub-layer Index | Viscosity (Pa s) | Temperature (K) | Ratio $z/d$ |
|---|---|---|---|
| 1 | $10^{14}$ | 273.0 | 1.00 |
| 2 | $10^{15}$ | 249.1 | 0.93 |
| 3 | $10^{16}$ | 229.0 | 0.86 |
| 4 | $10^{17}$ | 212.0 | 0.80 |
| 5 | $10^{18}$ | 197.3 | 0.75 |
| 6 | $10^{19}$ | 184.5 | 0.70 |
| 7 | $10^{20}$ | 173.3 | 0.65 |
| 8 | $10^{21}$ | 163.3 | 0.60 |
| 9 | $10^{22}$ | 154.5 | 0.56 |

## Appendix B. Example of libration calculations

Here we give an example of calculation using the methodology described in Section 2. In this example, the interior model has a shell thickness of 20.4 km. Table A2 gives the calculations.

**Table A2**. Example of libration calculation for the shell thickness of 20.4 km

| Physical quantity | Symbol | Value |
|---|---|---|
| Shell thickness (km) | d | 20.4 |
| Radial displacements at the upper surface of each layer | $y^i$ | 0.00426098 |
| | $y^o$ | 0.405993 |
| | $y^s$ | 0.394497 |
| Dynamical Love number of each layer | $k_2^i$ | 0.000143322 |
| | $k_2^{ob}$ | -0.000057987 |
| | $k_2^{ot}$ | 0.012261138 |
| | $k_2^s$ | 0.003988245 |
| Coefficients in equations (E5-E6) | $K_1$ | $3.05 \times 10^{20}$ |
| | $K_2$ | $-6.02 \times 10^{19}$ |
| | $K_3$ | $1.23 \times 10^{20}$ |
| | $K_4$ | $-5.88 \times 10^{19}$ |
| | $K_5$ | $1.44 \times 10^{20}$ |
| | $K_6$ | $4.28 \times 10^{19}$ |
| Polar moment of inertia for the shell and the core | $C_s$ | $5.36 \times 10^{29}$ |
| | $C_i$ | $1.03 \times 10^{30}$ |
| Total dynamical Love number | $k_2$ | 0.0163274 |
| Dissipation factor | $Q_s$ | 40.18 |
| Diurnal libration amplitude (km) | $\gamma_s \cdot R_s$ | 0.414 |

*Physical quantity uses the SI unit if its unit is not specified.